
\input harvmac
\input epsf
%
%
\newcount\figno
\figno=0
\def\fig#1#2#3{
\par\begingroup\parindent=0pt\leftskip=1cm\rightskip=1cm\parindent=0pt
\baselineskip=11pt \global\advance\figno by 1 \midinsert
\epsfxsize=#3 \centerline{\epsfbox{#2}} \vskip 12pt {\bf Fig.\
\the\figno: } #1\par
\endinsert\endgroup\par
}
\def\figlabel#1{\xdef#1{\the\figno}}
\overfullrule=0pt
\parskip=0pt plus 1pt
\sequentialequations

\def\half{{1\over 2}}
\def\del{\partial}
\def\CC{\hbox{\bf C}}

\def\Qp{\hbox{\bf Q}_p}
\def\ZZ{\hbox{\bf Z}}

\def\Re{\hbox{\bf R}}
\def\Zp{\hbox{\bf Z}_p}

\newcount\figno
\figno=0
\def\fig#1#2#3{
\par\begingroup\parindent=0pt\leftskip=1cm\rightskip=1cm\parindent=0pt
\baselineskip=11pt
\global\advance\figno by 1
\midinsert
\epsfxsize=#3
\centerline{\epsfbox{#2}}
\vskip 12pt
\centerline{Fig.\ \the\figno: #1}\par
\endinsert\endgroup\par
}
\def\figlabel#1{\xdef#1{\the\figno}}
\def\np{Nucl.\ Phys.}
\def\pl{Phys.\ Lett.}

\def\jhep{JHEP} 

\def\l{\left}
\def\r{\right}
\def\({\left(}
\def\){\right)}
\def\'{\prime}
\def\wt{\widetilde}
\def\to{\rightarrow}
\def\hf{{1 \over 2}}
\def\d{\partial}

\def\sitarel#1#2{\mathrel{\mathop{\kern0pt #1}\limits_{#2}}}

\def\inbar{\,\vrule height1.5ex width.4pt depth0pt}
\def\IB{\relax{\rm I\kern-.18em B}}
\def\IC{\relax\hbox{$\inbar\kern-.3em{\rm C}$}}
\def\ID{\relax{\rm I\kern-.18em D}}
\def\IE{\relax{\rm I\kern-.18em E}}
\def\IF{\relax{\rm I\kern-.18em F}}
\def\IG{\relax\hbox{$\inbar\kern-.3em{\rm G}$}}
\def\IH{\relax{\rm I\kern-.18em H}}
\def\II{\relax{\rm I\kern-.18em I}}
\def\IK{\relax{\rm I\kern-.18em K}}
\def\IL{\relax{\rm I\kern-.18em L}}
\def\IM{\relax{\rm I\kern-.18em M}}
\def\IN{\relax{\rm I\kern-.18em N}}
\def\IO{\relax\hbox{$\inbar\kern-.3em{\rm O}$}}
\def\IP{\relax{\rm I\kern-.18em P}}
\def\IQ{\relax\hbox{$\inbar\kern-.3em{\rm Q}$}}
\def\IR{\relax{\rm I\kern-.18em R}}
\font\cmss=cmss10 \font\cmsss=cmss10 at 7pt
\def\IZ{\relax\ifmmode\mathchoice
{\hbox{\cmss Z\kern-.4em Z}}{\hbox{\cmss Z\kern-.4em Z}}
{\lower.9pt\hbox{\cmsss Z\kern-.4em Z}}
{\lower1.2pt\hbox{\cmsss Z\kern-.4em Z}}\else{\cmss Z\kern-.4em Z}\fi}
\def\1{{1\hskip -3pt {\rm l}}}



\def\SL{{\rm SL}}

\def\m{\mu}
\def\n{\nu}
\def\ep{\epsilon}
\def\Z{{\bf Z}}

\def\x{\xi}
\def\t{\tau}
\def\Int{\int_{{\bf Q}_{p}}}
\def\IntZ{\int_{{\bf Z}_{p}}}

\def\abp#1{{|{#1}|_{p}}}
\def\abpsq#1{{|{#1}|^2_{p}}}
\def\sgn{{\rm sgn}_\tau}

\def\cD{{\cal D}}
\def\G{\Gamma_{\tau}}
\def\Gm{\Gamma}
\def\D{\Delta}
\def\a{\alpha}
\def\b{\beta}

\def\om{\omega}

\def\ch{\chi_{p}}
\def\wt{\widetilde}

\def\ve{\varepsilon}
\def\mod{{\rm mod~}}

%
%
\Title{\vbox{\baselineskip12pt\hbox{hep-th/0409311}%
\hbox{HRI-P-0408001}%
\hbox{UT-04-29}
}}%
{\vbox{
\centerline{Towards $p\,$-Adic String in Constant $B$-Field}
}}

{\vskip -20pt\baselineskip 14pt

\centerline{
Debashis Ghoshal\foot{On sabbatical leave from the Harish-Chandra 
Research Institute, Chhatnag Road, Allahabad 211 019, India.}
and Teruhiko Kawano} 

\bigskip
\centerline{\it Department of Physics, University of Tokyo}
\centerline{\it Hongo, Tokyo 113-0033, Japan}
\smallskip
\centerline{\tt ghoshal, kawano@hep-th.phys.s.u-tokyo.ac.jp}
\smallskip

\vglue .3cm

\bigskip\bigskip

\noindent Spacetime properties of the tachyon of the $p$-adic string
theory can be derived from a (non-local) action on the $p$-adic line
{\bf Q}$_p$, thought of as the boundary of the `worldsheet'. We show
that a term corresponding to the background of the antisymmetric
second rank tensor field $B$ can be added to this action. We examine
the consequences of this term, in particular, its relation to a
noncommutative deformation of the effective theory of the $p$-adic
tachyon in spacetime.

}
\Date{September, 2004}

\lref\SenRev{A.\ Sen, {\it Non-BPS states and branes in 
string theory}, [{\tt hep-th/9904207}].}

\lref\NCTach{K.\ Dasgupta, S.\ Mukhi and G.\ Rajesh, {\it 
Noncommutative tachyons}, \jhep\ {\bf 0006} (2000) 022 
[{\tt hep-th/0005006}];
\hfill\break
J\ Harvey, P.\ Kraus, F.\ Larsen and E.\ Martinec, {\it
D-branes and strings as non-commutative solitons}, \jhep\
{\bf 0007} (2000) [{\tt hep-th/0005031}]. }

\lref\FrOl{P.G.O.\ Freund and M.\ Olson,
{\it Nonarchimedean strings}, \pl\ {\bf B199} (1987) 186.}

\lref\FrWi{P.G.O.\ Freund and E.\ Witten, {\it Adelic string 
amplitudes}, \pl\ {\bf B199} (1987) 191.}

\lref\BFOW{L.\ Brekke, P.G.O.\ Freund, M.\ Olson and 
E.\ Witten, {\it Non-archimedean string dynamics},
Nucl.\ Phys.\  {\bf B302} (1988) 365.}

\lref\FrOka{
P.H.\ Frampton and Y.\ Okada,
{\it Effective scalar field theory of p-adic string},
Phys.\ Rev.\  {\bf D37} (1988) 3077. }

\lref\Fram{
P.H.\ Frampton and Y.\ Okada,
{\it The p-adic string N point function},
Phys.\ Rev.\ Lett.\  {\bf 60} (1988) 484\semi
P.H.\ Frampton, Y.\ Okada and M.R.\ Ubriaco,
{\it On adelic formulas for the p-adic string},
Phys.\ Lett.\  {\bf B213} (1988) 260.}

\lref\Spok{
B.L.\ Spokoiny,
{\it Quantum geometry of non-archimedean particles and strings},
Phys.\ Lett.\  {\bf B207} (1988) 401. }

\lref\Pari{
G.\ Parisi,
{\it On p-adic functional integrals},
Mod.\ Phys.\ Lett.\  {\bf A3} (1988) 639.}

\lref\Zha{
R.B.\ Zhang,
{\it Lagrangian formulation of open and closed p-adic strings},
Phys.\ Lett.\  {\bf B209} (1988) 229.}

\lref\Zabro{
A.V.~Zabrodin,
{\it Nonarchimedean String Action And Bruhat-Tits Trees},
Mod.\ Phys.\ Lett.\ A {\bf 4}, 367 (1989)\semi
A.V.\ Zabrodin,
{\it Non-archimedean strings and Bruhat-Tits trees},
Commun.\ Math.\ Phys.\  {\bf 123} (1989) 463.}

\lref\CMZ{
L.O.\ Chekhov, A.D.\ Mironov and A.V.\ Zabrodin, {\it
Multiloop calculations in p-adic string theory and Bruhat-Tits trees},
Commun.\ Math.\ Phys.\  {\bf 125} (1989) 675.}

\lref\FrNi{
P.H.\ Frampton and H.\ Nishino,
{\it Stability analysis of p-adic string solitons},
Phys.\ Lett.\  {\bf B242} (1990) 354.}

\lref\MaZa{
A.V.\ Marshakov and A.V.\ Zabrodin,
{\it New p-adic string amplitudes},
Mod.\ Phys.\ Lett.\  {\bf A5} (1990) 265.}

\lref\Chekhov{
L.O.\ Chekhov and Y.M.\ Zinoviev, {\it
p-Adic string compactified on a torus},
Commun.\ Math.\ Phys.\ {\bf 130} (1990) 130.}

\lref\BrFrRev{
L.\ Brekke and P.G.O.\ Freund, {\it p-Adic numbers in physics}, 
Phys.\ Rep.\ {\bf 133} (1993) 1, and references therein.}

\lref\GhSeP{
D.\ Ghoshal and A.\ Sen, {\it Tachyon condensation and brane descent
relations in $p$-adic string theory}, \np\ {\bf B584} (2000) 
300, [{\tt hep-th/0003278}].}

\lref\MinaWH{
J.A.~Minahan,
{\it Mode interactions of the tachyon condensate in p-adic string theory},
JHEP {\bf 0103}, 028 (2001), [{\tt hep-th/0102071}].}

\lref\MinaPD{
J.A.~Minahan,
{\it Quantum corrections in p-adic string theory}, {\tt hep-th/0105312}.
}

\lref\MZVX{
N.~Moeller and B.~Zwiebach,
{\it Dynamics with infinitely many time derivatives and rolling tachyons},
JHEP {\bf 0210}, 034 (2002), [{\tt hep-th/0207107}].
}

\lref\YangNM{
H-t.~Yang,
{\it Stress tensors in p-adic string theory and truncated OSFT},
JHEP {\bf 0211}, 007 (2002), [{\tt hep-th/0209197}].
}

\lref\VolovichZH{
Y.~Volovich, {\it Numerical study of nonlinear equations with infinite
number of derivatives}, J.\ Phys.\ A {\bf 36}, 8685 (2003), [{\tt
math-ph/0301028}].
}

\lref\MoellerGG{
N.~Moeller and M.~Schnabl,
{\it Tachyon condensation in open-closed p-adic string theory},
JHEP {\bf 0401}, 011 (2004), [{\tt hep-th/0304213}].
}

\lref\VlVo{
V.S.~Vladimirov and Y.I.~Volovich,
{\it On the nonlinear dynamical equation in the p-adic string theory},
Theor.\ Math.\ Phys.\  {\bf 138}, 297 (2004), [{\tt math-ph/0306018}].
}

\lref\GomisEN{
J.~Gomis, K.~Kamimura and T.~Ramirez,
{\it Physical degrees of freedom of non-local theories},
Nucl.\ Phys.\ B {\bf 696}, 263 (2004), [{\tt hep-th/0311184}].
}

\lref\Volo{
I.V.\ Volovich,
{\it p-Adic string},
Class.\ Quant.\ Grav.\  {\bf 4} (1987) L83; \hfill\break
B.\ Grossman,
{\it p-Adic strings, the Weyl conjectures and anomalies},
Phys.\ Lett.\  {\bf B197} (1987) 101.}

\lref\GeSh{
A.A.~Gerasimov and S.L.~Shatashvili,
{\it On exact tachyon potential in open string field theory},
JHEP {\bf 0010}, 034 (2000),
[{\tt hep-th/0009103}].
}

\lref\WiBSFT{
E.\ Witten,
{\it On background independent open string field theory},
Phys.\ Rev.\  {\bf D46}, 5467 (1992)
[{\tt hep-th/9208027}];\hfill\break
E.\ Witten, {\it
Some computations in background independent off-shell string theory},
Phys.\ Rev.\  {\bf D47}, 3405 (1993)
[{\tt hep-th/9210065}].}

\lref\ShBSFT{
S.L.\ Shatashvili, {\it
Comment on the background independent open string theory},
Phys.\ Lett.\  {\bf B311}, 83 (1993)
[{\tt hep-th/9303143}];\hfill\break
S.L.\ Shatashvili, {\it
On the problems with background independence in string theory},
{\tt hep-th/9311177}. }

\lref\KuMaMo{
D.~Kutasov, M.~Marino and G.~Moore,
{\it Some exact results on tachyon condensation in string field theory},
JHEP {\bf 0010}, 045 (2000)
[{\tt hep-th/0009148}].
}

\lref\GhSeN{
D.~Ghoshal and A.~Sen, {\it Normalisation of the background
independent open string field theory action}, JHEP {\bf 0011}, 021
(2000) [{\tt hep-th/0009191}]\semi
D.~Ghoshal,
{\it Normalization of the boundary superstring field theory action},
in Strings 2001, Eds.\ A.\ Dabholkar {\it et al}, AMS (2002)
[{\tt hep-th/0106231}].
}

\lref\Cornalba{
L.~Cornalba,
{\it Tachyon condensation in large magnetic fields with background 
independent string field theory},
Phys.\ Lett.\ B {\bf 504}, 55 (2001)
[arXiv:hep-th/0010021].
}

\lref\Okuyama{
K.~Okuyama,
{\it Noncommutative tachyon from background independent open string 
field theory},
Phys.\ Lett.\ B {\bf 499}, 167 (2001)
[arXiv:hep-th/0010028].
}

\lref\DouglasBA{
M.R.~Douglas and N.A.~Nekrasov,
{\it Noncommutative field theory},
Rev.\ Mod.\ Phys.\  {\bf 73}, 977 (2001)
[{\tt hep-th/0106048}].
}

\lref\HarveyYN{
J.~A.~Harvey,
{\it Komaba lectures on noncommutative solitons and D-branes},
{\tt hep-th/0102076}.
}

\lref\GoMiSt{
R.\ Gopakumar, S.\ Minwalla and A.\ Strominger,
{\it Noncommutative solitons}, \jhep\ {\bf 0005} (2000) 020.
[{\tt hep-th/0003160}].}

\lref\Schom{
V.~Schomerus,
{\it D-branes and deformation quantization},
JHEP {\bf 9906}, 030 (1999)
[{\tt hep-th/9903205}].
}

\lref\SW{
N.~Seiberg and E.~Witten,
{\it String theory and noncommutative geometry},
JHEP {\bf 9909}, 032 (1999)
[{\tt hep-th/9908142}].
}

\lref\moscow{
D.~Ghoshal, {\it Noncommutative p-tachyon}, to appear in the
Proceedings of {\it p-Adic MathPhys 2003}, Moscow. }

\lref\ncpadic{
D.~Ghoshal,
{\it Exact noncommutative solitons in p-adic strings and BSFT},
hep-th/0406259. }

\lref\GGPSP{
I.M.~Gelfand, M.I.~Graev and I.I.~Pitaetskii-Shapiro,
{\it Representation theory and automorphic functions}, Saunders (1969).}

\lref\KobP{
N.~Koblitz,
{\it p-Adic numbers, p-adic analysis and zeta functions}, 
GTM 58, Springer-Verlag (1977).}

\lref\RobP{
A.M.~Robert,
{\it A course in p-adic analysis}, GTM 198, Springer-Verlag (2000).}

\lref\RussP{
V.S.~Vladimirov, I.V.~Volovich and E.I.~Zelenov,
{\it p-Adic Analysis and Mathematical Physics},
(Series on Soviet and East European Mathematics, Vol.1) World Scientific, 
1994. }

\lref\FTACNY{
E.S.~Fradkin and A.A.~Tseytlin,
{\it Nonlinear electrodynamics from quantized strings},
Phys.\ Lett.\ B {\bf 163}, 123 (1985)\semi
A.~Abouelsaood, C.G.~Callan, C.R.~Nappi and S.A.~Yost,
{\it Open strings in background gauge fields},
Nucl.\ Phys.\ B {\bf 280}, 599 (1987).
}

\lref\MelzerHE{
E.~Melzer,
{\it Nonarchimedean conformal field theories},
Int.\ J.\ Mod.\ Phys.\ A {\bf 4}, 4877 (1989).
}

\lref\Heaviside{
I.~Y.~Arefeva, B.~G.~Dragovic and I.~V.~Volovich,
{\it Open and closed p-adic strings and quadratic extensions of 
number fields},
Phys.\ Lett.\ B {\bf 212}, 283 (1988).
}

\lref\Fairlie{
D.~B.~Fairlie and K.~Jones,
{\it Integral representations for the complete four- and five-point 
Veneziano amplitudes},
Nucl.\ Phys.\ B {\bf 15}, 323 (1970).
}

\lref\Barnett{
A.~Barnett,
{\it p-Adic amplitudes}, IMPERIAL-TP-90-91-02.
}

\lref\NeVa{
S.~Nechaev and O.~Vasilyev,
{\it On metric structure of ultrametric spaces}, 
J.\ Phys.\ A {\bf 37}, 3783 (2004).}

\lref\Grange{
P.~Grange,
{\it Deformation of p-adic string amplitudes in a magnetic field},
hep-th/0409305.}

{\nopagenumbers

\ftno=0
}

\newsec{Introduction}
The tachyon sector of the D-branes of the open $p$-adic string theory
is amenable to exact analytic calculation. It turns out that in many
ways these theories behave like the ordinary bosonic string. The
history of the $p$-adic string is also not unlike the ordinary string
theory. It was introduced in Ref.\FrOl, which proposed the
Koba-Nielsen formula for the tree amplitude of $N$ tachyons of the
$p$-adic theory by generalizing the formula from the real numbers to
the $p$-adic numbers. Namely, the integral was taken over (a quadratic
extension of) the $p$-adic number field $\Qp$ and the absolute values
in the integrands were replaced by the non-archimedian norm.  It was
shown that this prescription defines a consistent string theory.  Soon
afterwards it was realized\refs{\FrWi\BFOW\FrOka-\Fram} that these
integrals can be computed exactly. The spacetime effective field
theory of the tachyons is, therefore, known {\it exactly}. This was
written down in Refs.\refs{\BFOW,\FrOka}. (Let us notice that in this
approach, spacetime is the usual Minkowski one. A more exotic type of
$p$-adic string living in $p$-adic spacetime was studied in \Volo, but
we will not consider it here.)

Although the prescription to define the $p$-adic string was well
motivated mathematically and paid rich dividends, it did not shed much
light on the nature of the $p$-adic string itself. The development in
this direction came from the works of
\refs{\Spok\Pari\Zha\Zabro-\CMZ}.  In Refs.\refs{\Spok\Pari-\Zha}, a
non-local action on the $p$-adic field $\Qp$ was proposed for
$X^\mu(\x)$ ($\x\in\Qp$), the target space coordinates.  Finally, a
`worldsheet' theory was given in \refs{\Zabro,\CMZ}. The `bulk' of the
open string worldsheet turns out to be an infinite Bethe lattice, also
called a Bruhat-Tits tree, with coordination number $p+1$; the
boundary of which is isomorphic to $\Qp$. The Polyakov action on the
`worldsheet' is the usual lattice action for the free massless fields
$X^\mu$. It was shown in \Zabro\ that starting with a finite lattice
and inserting the tachyon vertex operators on the boundary, one
recovers the prescription of
\refs{\FrOl\FrWi\BFOW\FrOka-\Fram} in the thermodynamic limit. For
related works on the $p$-adic string theory, see
\refs{\FrNi\MaZa-\Chekhov}, as well as the review \BrFrRev.

More recently, the $p$-adic string theory have come to focus through
the realization that the exact spacetime theory of its tachyon allows
one to study nonperturbative aspects of string dynamics, like the
process of tachyon condensation. In \GhSeP, the solitons of the
effective theory of the $p$-adic tachyon\refs{\BFOW,\FrOka,\FrNi} were
identified with the D-branes and shown that the dynamics is according
to the behaviour conjectured by Sen\SenRev. Moreover, it turns out
that in the $p\to 1$ limit\foot{See also the prescient comments in
\Spok.}\GeSh, the theory provides an approximation to the boundary
string field theory (BSFT)\refs{\WiBSFT,\ShBSFT} of ordinary strings,
the formalism of which was useful in proving the Sen
conjectures\refs{\GeSh,\KuMaMo,\GhSeN}.  Various properties of the
$p$-adic string have been explored recently in
\refs{\MinaWH\MinaPD\MZVX\YangNM\VolovichZH\MoellerGG\VlVo-\GomisEN}.

Despite these progress, much remains to be understood about the
$p$-adic strings. We know some properties of its D-branes in flat
space, though restricted mostly to the tachyon sector. Recently 
a first step towards understanding the behaviour of $p$-adic strings
in a non-trivial background was taken in \refs{\moscow,\ncpadic}. The
motivation comes from the fact that a constant value of the second
rank antisymmetric tensor field $B$ in ordinary string theory has the
effect of providing a noncommutative deformation of the open string
fields in the target space. Thus the effective spacetime theory of the
$p$-adic tachyon was deformed by introducing a noncommutative
parameter $\theta$. Gaussian solitons corresponding to D-branes were
obtained for all values of $\theta$ and shown to interpolate smoothly
from the $p$-adic soliton\refs{\BFOW} to the noncommutative GMS
soliton\GoMiSt. It was shown that this continues to be the case down
to $p\to 1$, {\it i.e.}, in the BSFT of the ordinary string theory;
thus providing a new insight to the effectiveness of the approach of
Refs.\NCTach\ in understanding tachyon dynamics.

In this paper, we would like to examine a possible `worldsheet' origin
of the noncommutativity introduced in \refs{\moscow,\ncpadic} to the
effective action of the $p$-adic tachyon. We will modify the nonlocal
action\refs{\Spok\Pari-\Zha} on the {\it boundary} of the `worldsheet'
in analogy with the ordinary string theory in a constant
$B$-field\foot{It turns out that this can be done consistently for a
subset of primes $p$, namely $p=3$ (mod 4), which roughly are half of
all the prime numbers. However, once we are in the domain of the
effective theory in spacetime, there is no reason why the results
cannot be continued beyond this set, indeed to all integers.}. We show
that the correlation functions with any number of the tachyon vertex
operators can be solved in the presence of the coupling to the
$B$-field (Sec.~3).  Unfortunately, however, the deformation does not
respect the complete GL(2,$\Qp$) symmetry---it is only invariant under
the infinitesimal transformation---the underlying reason being the
lack of a natural order in $\Qp$. As a result, the four-tachyon
scattering amplitude does not quite agree with that computed from the
(deformed) effective field theory (Sec.~4). However, we discuss some
formal `time ordered' Green functions and related commutators of the
tachyon field. As an alternative approach, we explore the possibilies
of deforming the Koba-Nielsen amplitudes (Sec.~5) and end with some
comments (Sec.~6).  For completeness, the `worldsheet' approach to the
$p$-adic string, developed in Refs.\Zabro\ and \refs{\Spok-\Zha}, in
flat spacetime with no background field, is briefly recapitulated
(Sec.~1). Two appendices provide a collection of results from $p$-adic
analysis useful for our purpose (Appendex A); and a brief derivation
of an infrared regulated integral (Appendix B).


\newsec{`Worldsheet' action of the $p$-adic string in the trivial 
background} 
In this section we will review the `worldsheet' approach
to the $p$-adic open string in flat spacetime with no other background
field, as proposed in Refs.\refs{\Spok\Pari\Zha\Zabro-\CMZ}.  

Let us start by recalling the `worldsheet' construction of the
$p$-adic string given in Ref.\Zabro. The interior of the worldsheet,
analogous to the unit disc or the upper half-plane of the usual theory
is an infinite lattice with no closed loops, {\it i.e.}, a tree ${\cal
T}_p$ in which $p+1$ edges meet at each vertex (see Fig.~1). This is
known to mathematicians as the Bruhat-Tits tree and is familiar to
physicists as the Bethe lattice. The boundary of the tree ${\cal
T}_p$, defined as the union of all infinitely remote vertices, can be
identified with the $p$-adic field $\Qp$. On the other hand, the tree
${\cal T}_p$ is the (discrete) homogeneous space
PGL(2,$\Qp$)/PGL(2,$\Zp$): the coset obtained by modding PGL(2,$\Qp$)
by its maximal compact subgroup\foot{This construction parallels the
case of the usual string theory, in which the UHP is the homogeneous
coset PSL(2,$\Re$) modulo its maximal compact subgroup SO(2).}
PGL(2,$\Zp$). The action of PGL(2,$\Qp$) on $\Qp$ extends naturally to
${\cal T}_p$.

\fig{A finite part of the `worldsheet' of the $3$-adic string: the 
tree ${\cal T}_p$ for $p=3$.}{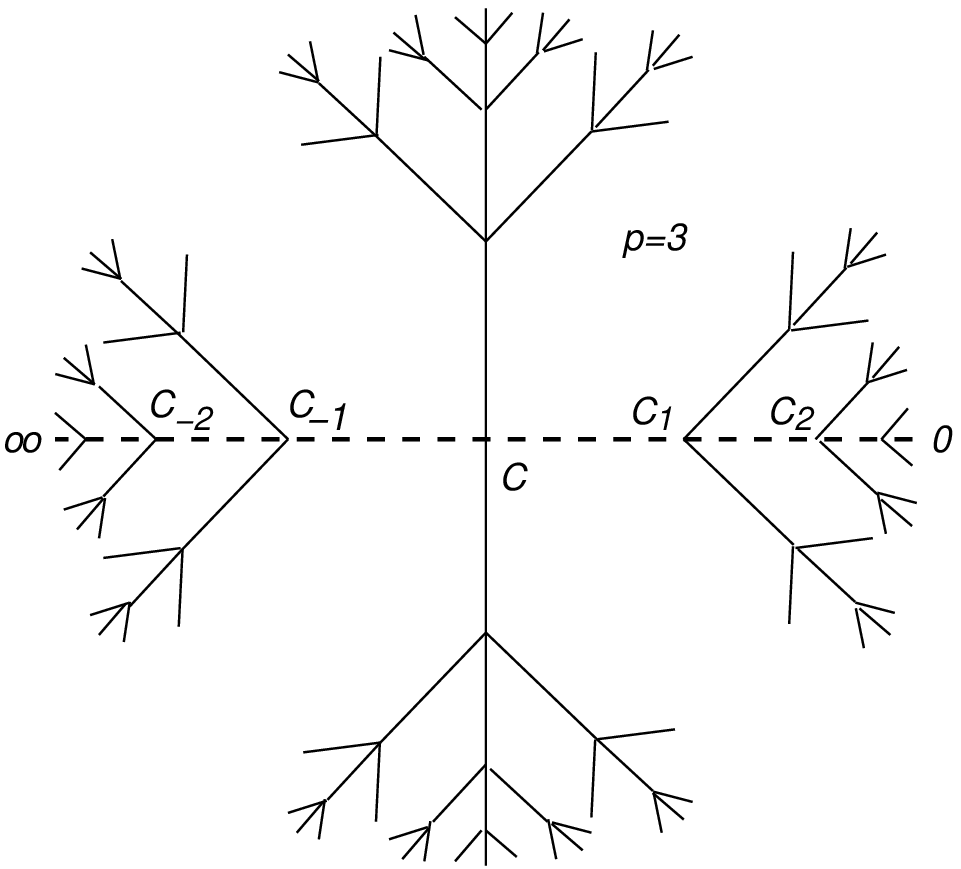}{2.5in}

Let $z$ label the vertices of ${\cal T}_p$ and $e$ its
edges. We will use $\xi$ to label the points on the boundary
$\del{\cal T}_p\equiv\Qp$. The spacetime coordinates $X^\mu(z)$ are
functions on the lattice. The difference $\Delta_eX^\mu=X^\mu(z')-
X^\mu(z)$, where $z$ and $z'$ are the end-points of the edge $e$, is
the finite lattice analogue of the directional derivative. Likewise,
the Polyakov action on the $p$-adic worldhsheet is the free action
\eqn\trivwsaction{
S_p[X] = \half\beta_p\,\sum_{\{e\}}\left(
\Delta_e X^\mu(z)\right)^2.}
The variation of the above gives rise to
\eqn\varaction{
\delta S_p = -\beta_p\,\left[\sum_{\{z\}}\delta X^\mu(z)
\nabla^2 X^\mu(z) + (p-1)\int_{\del{\cal T}_p}d\mu_\xi\, 
\delta X^\mu(\xi)\,{\cal D}_n^{(p)}X^\mu(\xi)\right] , }
where
\eqn\laplace{
\nabla^2 X^\mu(z) = \sum_{i=1}^{p+1}X^\mu(z_i) - (p+1)X^\mu(z), }
($z_i$ are the $p+1$ nearest neighbours of $z$), is the lattice
Laplacian, ${\cal D}_n^{(p)}X^\mu(\xi)$ is an appropriately defined
normal derivative\foot{This is defined in terms of the limit of
difference of $f(z)$ and $f(\xi)$ as the interior point $z$ of the
tree ${\cal T}_p$ approach the boundary $\Qp=\del{\cal T}_p$.} at the
boundary point $\xi$ and ${d\mu}_\xi$ is a measure on the boundary
${\cal T}_p$ (see \Zabro\ for the precise definitions). The tension of
the $p$-adic string is $\beta_p$.  We see that either the Neumann
(${\cal D}_n^{(p)}X^\nu(\xi)=0$) or the Dirichlet 
($X^\nu(\xi)=\,$constant) boundary condition can be imposed on a
spacetime coordinate $X^\nu$. The solutions of the classical equations
of motion $\nabla^2 X^\mu(z)=0$ are harmonic functions on the tree
${\cal T}_p$.

We are interested in the scattering amplitude of $N$ tachyonic
scalars.  This requires us to compute the correlation function of the
$N$ vertex operators $\exp\left(ik^I\cdot{X}\right)$, $(k^I)^2=2$,
corresponding to the tachyon with momentum $k^{I}\; (I=1,\cdots,N)$,
inserted at the boundary points $\xi_I\in\del{\cal T}_p=\Qp$. To this
end, we work with a finite part of the lattice of radius $R$ centred
at the point $C$ (see Fig.~1) and define the scattering amplitide
${\cal A}_p^{(N)}$ as the limit $R\to\infty$ of the finite lattice
correlator:
\eqn\VOcorr{
{\cal A}_p^{(N)} = \lim_{R\to\infty}{1\over Z_p}\int {\cal D}X\,
\exp\left(-S_p[X] + i\sum_{I=1}^N k^I\cdot{X}(\xi_I)\right), }
where $\xi_I$ are points on the boundary at radius $R$. The measure
${\cal D}X=\prod_{\mu,z} dX^\mu(z)$ is the usual measure and the
normalization in \VOcorr\ is by the partition function
\eqn\PF{
Z_p = \int {\cal D}X\,e^{-S_p[X]}. }
The regularity of the thermodynamic limit $R\to\infty$ determines the
string tension to be $\beta_p=1/\ln p$, which also happens to be the
condition for the projective GL(2,$\Qp$) invariance of the tachyon
scattering amplitude. In Ref.\Zabro, it is shown that the
expression \VOcorr\ is the $p$-adic analogue of the Koba-Nielsen
amplitude
\eqn\pKN{
{\cal A}_p^{(N)} = \Int\prod_{I=1}^{N-3}d\xi_I\,
|\xi_I|_p^{k^I\cdot{k}^N} |1-\xi_I|_p^{k^I\cdot{k}^{N-1}}
\prod_{1\le I< J\le N-3}|\xi_I-\xi_J|_p^{k^I\cdot{k}^J}, }
where the projective invariance is used to fix the boundary points
$\xi_{N-2}=0$, $\xi_{N-1}=1$ and $\xi_N=\infty$.

It is also possible to integrate over the degrees of freedom in the 
interior of the tree ${\cal T}_p$ to arrive at a non-local action
on the boundary (see Appendix C of Ref.\Zabro\ for details):
\eqn\bndryS{
{\cal S}_p = {p(p-1)\beta_p\over 4(p+1)}\,\Int d\xi\,
d\xi'\,{\left(X^\mu(\xi)-X^\mu(\xi')\right)^2\over |\xi-\xi'|_p^2}. }
This action was proposed in \refs{\Spok-\Zha} and shown to yield the
correlators of the tachyon vertex operators inserted on the boundary.
In the following, this will be our starting point for coupling the
$B$-field.  Let us note that in the case of the usual bosonic string
theory too, there is an analogue of this non-local action---it is the
one obtained by the obvious substitutions in Eq.\bndryS.


\newsec{A constant $B$-field background in the boundary effective 
action}

In ordinary string theory, a constant background $B$-field contributes
a total derivative in the worldsheet action. Hence its only effect is in
the boundary 
\eqn\Bbndryterm{
\hf\int_{\Sigma}d^2z\,\varepsilon^{\a\b}
\del_\a X^\m(z)\del_\b X^\n(z)B_{\m\n}
=\int_{\del\Sigma}d\xi\, B_{\m\n}X^\m(\xi)\del_t X^\n(\xi),}
where $\del_t$ is the tangential derivative along the boundary
$\d\Sigma$, which we label by $\xi$. (Formally, this is the same as
the insertion of the vertex operator for a background gauge field
$A_\m\l[X(\t)\r]\sim{B}_{\m\n}X^\n(\t)$.) In principle therefore, it is
a straightforward problem to generalize the effect of a constant
$B$-field in the $p$-adic string theory: add the $p$-adic analogue of
the boundary term \Bbndryterm\ to the action \bndryS. However, one is
immediately faced with a problem: there is no natural notion of a
(tangential) derivative along the $p$-adic line $\Qp$ (see
\refs{\GGPSP\KobP\RobP-\RussP} for aspects of $p$-adic analysis). 

The same problem was encountered by the authors of \MaZa\ in their
computation of the correlation function involving the vertex operator
of a vector field $\epsilon_\m(k)\,\del_t{X}^\m e^{ik\cdot X}(\xi)$ in
the $p$-adic string theory. The solution proposed there is to use the
Cauchy-Riemann relations for a pair of harmonic functions in the usual
case, and write the tangential derivative as\refs{\MaZa,\GGPSP}
\eqn\tangent{
\del^{(p)}_tX^\m(\xi) \equiv \del_\xi X^\mu
=\Int d\xi'\,{\sgn(\xi-\xi')\over\abpsq{\x-\x'}}\, X^\m(\xi'),}
where $\sgn{(\xi)}$ is an analogue of the sign function over $\Re$.
More precisely, it is a so called {\it multiplicative branching
character} on $\Qp$ correspoding to the quadratic extension\foot{The
subscript $\tau\in\Qp$ refers to a $p$-adic number such that
$\sqrt{\tau}\notin\Qp$. One obtains a quadratic extension
$\Qp(\sqrt{\tau})$ of the $p$-adic field by adjoining $\sqrt{\tau}$;
{\it i.e.}, by considering elements of the form $\xi+\sqrt{\tau}\eta$
for $\xi,\eta\in\Qp$. Unlike $\Re$ the quadratic extension of $\Qp$ is
not unique. In other words, there are several choices for $\tau$ and
the function $\sgn$ depends on this choice. It is worth emphasizing,
however, that $\sgn$ is a (real valued) function on $\Qp$, and does
not require that we extend it.} $\Qp(\sqrt{\tau})$ by $\sqrt{\tau}$.
It turns out that, up to equivalence, there are three choices
$\tau=\varepsilon$, $p$ and $\varepsilon p$ (where $\varepsilon$ is a
$(p-1)$-th root of unity). However, if we demand the antisymmetry
property of the sign function, as in $\Re$:
\eqn\antisymm{
\sgn{(-\xi)} = -\,\sgn{(\xi)}. }
holds only for the last two choices (namely, $\tau=p,\,\varepsilon p$)
with an {\it additional restriction} on the value of the prime $p=3$ 
(mod 4), which is satisfied by roughly half the prime numbers\GGPSP. 
The antisymmetry property \antisymm\ is of importance to us, so we will
restrict to these values of $p$. For a $\xi\in\Qp$, $\sgn(\xi)$ is 
defined as follows:
\eqn\defsgn{
\sgn{(\xi)}=\cases{+1, &if $\xi=\zeta^2_1-\tau\zeta_2^2$ for some
$\zeta_1,\zeta_2\in\Qp$,\cr
-1 &otherwise.} }
Let us recall that the function $\sgn$ was used in defining $p$-adic
string amplitudes with Chan-Paton 
factors\refs{\BFOW,\BrFrRev,\Heaviside}.

We are now in a position to put forward our proposal for the boundary
effective action in the presence of a constant background
$B$-field. It is:
\eqn\Bpaction{
\eqalign{
{\cal S}_p(B) &= {T_0\over2}\bigg[\Int \eta_{\m\n}
{\(X^\m(\xi)-X^\m(\xi')\)\(X^\n(\xi)-X^\n(\xi')\)\over
\abpsq{\xi-\xi'}}d\xi d\xi'\cr
&\qquad +i {1+p\over p^2\Gamma_\tau(-1)}\Int B_{\m\n}\,X^\m(\xi)
{\sgn(\xi-\xi')\over\abpsq{\xi-\xi'}}X^\n(\xi')d\xi d\xi'\bigg],
}}
where $\Gamma_\tau(\xi)$ is a generalized gamma function over the
$p$-adic field\foot{The generalized gamma functions $\Gamma_\tau(s)$
associated with $\sgn$, is not singular at zero or negative integer
arguments, or indeed anywhere on the complex $s$-plane.} discussed 
in Appendix A and 
\eqn\tension{
T_0 = {p(p-1)\over 2(p+1)\ln p}\,{1\over\alpha'} }
is the `string tension'. The antisymmetry of the $B$-field requires
that at least two directions are involved, thus the rank of $B$ is
always even. In the following, although we will not be precise about
it: if a (spatial) coordinate $X^\mu$ is involved, it is assumed that
$B$ has a component in that direction.

The normal derivative\refs{\Spok-\Zabro} on $\Qp$
\eqn\normal{
\del_n^{(p)}f(\xi) = \Int d\xi'\;{f(\xi')-f(\xi)\over
\abpsq{\xi'-\xi}} }
appears in the first line of Eq.\Bpaction. Let us note in parenthesis
that a normal derivative ${\cal D}_n^{(p)}$ was introduced in
Eq.\varaction. Neither the Haar mesaure $d\xi$ nor the normal
derivative \normal, match the corresponsing quantities in
Eq.\varaction.  However, the combination $d\xi\,\del_n^{(p)}f(\xi)$
equals ${d\mu}_\xi{\cal D}_n^{(p)}f$ (see \Zabro\ for details).

\subsec{Correlators in the presence of $B$-field}

In the rest of the section, we will analyze various consequences of
action \Bpaction\ using standard field theoretic techniques. We recall
that the $N$-point correlation function of the tachyon vertex operator
$e^{ik\cdot X}(\xi)$ in a constant $B$-field background is given by
the correlator
\eqn\Ntachcorr{
\left\langle e^{ik_1\cdot{X}}\l(\x_1\r)\,\cdots\, 
e^{ik_N\cdot{X}}\l(\x_N\r) \right\rangle_B
={\displaystyle\int{\cal D}X\, \exp\l(-{\cal S}_p(B)+
i\displaystyle\sum^N_{I=1}k^I\cdot{X}(\x_I)\r)
\over\displaystyle\int{\cal D}X\,\exp\l(-{\cal S}_p(B)\r)}. }
Let us introduce the generating function 
\eqn\genfn{
{\cal Z}\l[J\r]=\int{\cal D}X\; \exp\l[-{\cal S}_p(B)+i\Int 
d\xi\, J\cdot X(\xi)\r], }
which, when the external current $J_\mu(\xi)=\sum^N_{I=1}k_\mu^I\,
\delta\(\xi-\xi_I\)$, yields the $N$-point correlation function
\Ntachcorr. The generating function ${\cal Z}\l[J\r]$ satisfies the
Dyson-Schwinger equation 
\eqn\DysonS{
\Int d\xi'\,\D_{\mu\nu}(\xi-\xi'){\delta\ln{\cal Z}[J]\over
\delta{J}_\nu(\xi')}=-\,J_\mu(\xi), } 
where the operator $\D_{\m\n}$
\eqn\kineticop{
\D_{\m\n} = T_0 \l[ -\eta_{\mu\nu} \del_n^{(p)} + i {1+p\over p^2
\Gamma_\tau(-1)}B_{\m\n}\del_t^{(p)}\r]}
is a combination of the normal derivative $\del_n^{(p)}$ and
tanegntial derivative $\del_t^{(p)}$ defined in Eqs.\normal\ and
\tangent\ respectively. 

It remains to solve the Green function ${\cal G}^{\mu\nu}\l(\xi-\xi'\r)$ 
which is a kernel to the differential operator \kineticop
\eqn\Green{
\Int d\xi''\D_{\m\lambda}(\xi-\xi'')\,{\cal G}^{\lambda\nu}(\x''-\x')
=\delta_\mu^\nu\,\delta\l(\xi-\xi'\r). }
To this end, we solve the `equation of motion':
\eqn\eomXJ{
\Int d\xi'\,\D_{\m\n}\l(\xi-\xi'\r)\,X^\nu\l(\xi'\r) = -
J_\mu\l(\xi\r), }
by taking a Fourier transformation on the $p$-adic number $\Qp$, 
as we do for ordinary strings (See Appendix A and 
\refs{\GGPSP\KobP\RobP-\RussP} for details). In the Fourier space 
Eq.\eomXJ\ takes the form
\eqn\eomFourier{
T_0 {1+p\over{p}^2}
\big(\eta_{\mu\nu}- i B_{\mu\nu} \sgn\l(\omega\r)\big)\,
\abp{\omega}\,\wt{X}^\nu\l(\omega\r)
= - \wt{J}_\mu\l(\om\r). }
Introducing the open string metric $G_{\mu\nu}$ and the theta 
parameter $\theta^{\mu\nu}$ by the relation 
\eqn\Gtheta{
G^{-1}+{i\over2}\,{p-1\over\alpha' p\ln{p}\,\Gamma_\tau(0)}\,
\theta=\l({1\over \eta-iB}\r), }
we find that 
\eqn\XFourier{
\wt{X}^\mu\l(\omega\r) = - 
{p^2\over T_0(1+p)}\l(G^{\mu\nu}+{i\over2}\,{p-1\over\alpha'{p}
\ln{p}\,\Gamma_\tau(0)}\theta^{\mu\nu} \sgn\l({\omega}\r)\r)
{1\over\abp{\omega}}\sum^N_{I=1}k^I_\nu\ch\l(-\omega\x_I\r), }
where $\ch(\omega)$ is the $p$-adic analogue of the function
$e^{i\omega}$ (see Appendix A). We have also used the expression of
the external current $J_\mu=\sum^N_{I=1}k_\mu^I\,\delta\(\xi-\xi_I\)$,
appropriate for \Ntachcorr. Finally, after an inverse Fourier
transform, we arrive at the desired expression
\eqn\solnX{
X^\mu\l(\xi\r) = \sum^N_{I=1}k_\nu^I\left(\alpha' G^{\mu\nu}
\ln\abp{\x-\x_I} - {i\over2} \theta^{\mu\nu}\sgn\l(\xi-\x_I\r)
\right), }
To be precise, the expression above is obtained after an `infrared
regularization'. It is similar to the usual string theory and the
details are given in Appendix B.
 
The Green's function is therefore
\eqn\Greensol{
{\cal G}^{\mu\nu}\l(\xi-\xi'\r) =-\a'G^{\mu\nu}\ln\abp{\xi-\xi'}
+ {i\over2} \theta^{\mu\nu}\sgn\l(\xi-\xi'\r), } 
in terms of which we have a complete solution
\eqn\Zsol{
{\cal Z}[J]=\exp\l(-\half\displaystyle\sum^N_{I,J=1} 
k^I_\mu k^J_\nu {\cal G}^{\mu\nu}\l(\xi_I-\xi_J\r)\r) }
up to an irrelevant constant of integration. In particular, the
$N$-point correlation function \Ntachcorr\ is:
\eqn\Btachcorr{
\left\langle e^{ik^1\cdot{X}}(\xi_1)\cdots e^{ik^N\cdot{X}}(\xi_N)
\right\rangle_B = 
\sitarel{\prod^N_{I,J=1}}{I<J}
{\exp\left(-{i\over2}\theta^{\mu\nu}k^I_\mu k^J_\nu\sgn(\xi_I-
\xi_J)\right)\over|\xi_I-\xi_J|_p^{-\alpha' G^{\mu\nu}k^I_\mu 
k^J_\nu}}, }
where we have omitted the momentum conserving delta function.
This is the $p$-adic analogue of the result in Ref.\FTACNY\ for the
usual bosonic string and has the same form as the latter. 

\subsec{Formal consequences of the $B$-field}

Notwithstanding its apparently nice expression and the similarity with
the real case, there are problems with \Btachcorr. It is not invariant
under projective invariance of $\Qp$, {\it i.e.}, under GL(2,$\Qp$)
transformations. Looking back, we find that this is indeed a problem
with \Bpaction. As a matter of fact, even in the real case, the
amplitudes are invariant under those SL(2,$\Re$) transformations which
preserve the cyclic ordering of the tachyon vertex operators.  It is
here that we face a fundamental problem: one cannot define an order in
$\Qp$ that is compatible with its algebraic properties. The function
$\sgn$ is therefore not associated with any ordering. The best we can
do is to prove invariance under infinitesimal $\SL(2,\Qp)$
transformation. However, before we sketch the proof, let us provide
some formal argument, which show that the $B$-field background gives
rise to spacetime noncommutativity, at least formally.

We start by defining an ordering (`time ordering') in $\Qp$. This can
done by ordering the coefficients in the power series expansion of a
$p$-adic number: $\xi=p^N\left(\xi_0+\xi_1 p+\cdots\right)$ (given in
Appendix A), but let us add a further ingredient to it. The first
coefficient $\xi_0$ can be any of the non-zero elements of the finite
field ${\bf F}_p= \Zp/p\Zp\simeq \Z/p\Z$. Moreover, a primitive
$(p-1)$-th root of unity exists\GGPSP\ in ${\bf F}_p$. In other words,
there exists $\eta\in{\bf F}^*_p$, such that $\eta^{p-1}=1$ and powers
of $\eta$ generates ${\bf F}^*_p$. This implies that out of the
$(p-1)$ values of $\xi_0$, exactly half are odd/even powers of
$\eta$. It is now possible to show that for a fixed $N$, {\it i.e.},
for $|\xi|_p=p^{-N}$, exactly half have $\sgn(\xi)=+1$ and the other
half have $\sgn(\xi)=+1$. In terms of the `worldsheet' in Fig.~1, at
each node along the dashed line from zero to infinity, $(p-1)$
branches labelled by $\xi_0$ originate. Of these, $(p-1)/2$ correspond
to a $p$-adic number $\xi$ with $\sgn(\xi)=+1$ (respectively $-1$). By
convention, we can say that the positive (negative) set is the other
above (below) the dashed line, and each half can be ordered by the
norm and the coefficients in the power series. The purpose of this
exercise is to show that we can approach $\xi=0$ through a sequence
of points $\xi_I$ such that $|\xi_I|_p\to 0$ and $\sgn(\xi_I)$ fixed
to be either of $\pm 1$, exactly as in the case of the real string. 

We can now follow standard procedure\refs{\Schom,\SW} to write
\eqn\pcommutator{
\eqalign{
[X^\mu(0),X^\nu(0)] &= T\left(\left\langle X^\mu(0)X^\nu(0-)
\right\rangle - \left\langle X^\mu(0)X^\nu(0+)\right\rangle\right)\cr
&= \lim_{{\sgn(\xi)=-1\atop|\xi|_p\to 0}} \left\langle X^\mu(0)
X^\nu(\xi)\right\rangle\; - \lim_{\sgn(\xi)=+1\atop|\xi|_p\to 0}
\left\langle X^\mu(0)X^\nu(\xi)\right\rangle\cr
&=\; i\theta^{\mu\nu}, }}
where, we have made use of the Green function \Greensol. It is
possible to rephrase it in another way. The exact expression for the
correlators \Btachcorr, can be written equivalently as the formal
operator product expansion between the tachyon vertex operators on
$\Qp$:
\eqn\BtachOPE{
:e^{ik\cdot X}(\xi):\,:e^{ik'\cdot X}(\xi'):\; =\;
{\exp\left(-{i\over2}k_\mu k'_\nu\theta^{\mu\nu}\sgn(\xi-\xi')\right)
\over |\xi-\xi'|_p^{-\alpha' G^{\mu\nu}k^\mu {k'}^\nu}}
:e^{i(k+k')\cdot X}(\xi): + \cdots. }
Let us introduce, following Ref.\Schom,
\eqn\fieldinST{
\Phi\left(X(\xi)\right) = {1\over (2\pi)^{D}}\int_{{\bf R}^D}
d^Dk\,\tilde\Phi(k)\,e^{ik\cdot X}(\xi), }
where $\tilde\Phi(k)$ is the Fourier transform of the function
$\Phi(x)$. Notice that the object \fieldinST\ is similar to a string
field. More precisely, in the usual string theory in which $\xi$ labels 
the real line, $\phi(x)=\lim_{\xi\to 0}\Phi\left(X(\xi)\right)|0\rangle$
is the string field for the tachyon. Remaining in the real case, we recall
that the operator product expansion of the vertex operators is equivalent
to the statement about multiplication of the objects in \fieldinST. 
Without a $B$-field, $\Phi\left(X(1)\right)\Psi\left(X(0)\right)=
\left(\Phi\Psi\right)\left(X(0)\right)+\cdots$; this gets deformed to
a noncommutative product $\Phi\star\Psi$ in the presence of a constant
$B$-field. 

Returning to the case of the $p$-adic string, we will evaluate the product 
$\Phi\left(X(1)\right)\Psi\left(X(0)\right)$.  (Parenthetically, evaluating 
this with $\xi=1$ and $\xi'=0$ assumes that the GL(2,$\Qp$) invariance can 
be used to find a more general expression\MelzerHE.)
\eqn\PhiPsi{
\eqalign{
\Phi\left(X(1)\right)\Psi\left(X(0)\right) &=
{1\over (2\pi)^{D}}\int_{{\bf R}^D}
d^Dk\,\widetilde{(\Phi\star\Psi)}(k)\,e^{ik\cdot X}(0),\cr
\widetilde{(\Phi\star\Psi)}(k) &= 
{1\over (2\pi)^{D}}\int_{{\bf R}^D}d^Dk\,\tilde\Phi(k')
\tilde\Psi(k-k')\,e^{-{i\over2}k_\mu k'_\nu\theta^{\mu\nu}}. }}
Thus, it seems that as a result of the $B$-field in \Bpaction,
ordinary pointwise multiplication of functions of spacetime is
deformed to the noncommutative Moyal product.

It would, however, be prudent to interpret the above results with some
caution. The arguments we have used are rather formal. The $B$-field
does not respect GL(2,$\Qp$) invariance. And the order that we have
defined is not compatible with the algebraic properties of the field
$\Qp$. In particular, it is not invariant under the projective
symmetry. In order to demonstrate that the $B$-field leads to
spacetime noncommutativity, one needs to calculate all the tachyon
$N$-point functions, in the presence of the constant $B$-field. We
will discuss the difficulties with this in the next section.  

\subsec{Infinitesimal projective invariance}

Let us study the $p$-adic analog of the projective (M\"obius) 
transformation GL(2,$\Qp$) on the correlation functions \Btachcorr. 
The coordinates $\x_I$ $(I=1,\cdots,N)$ of the vertex operators are 
transformed as 
\eqn\Moebius{
\xi_I ~\to~ \xi'_I = {a\xi_I+b\over c\xi_I+d},\qquad ad-bc\ne 0,
}
where $a,b,c,d\in\Qp$. 
It is easy to see that the integrated $N$-point correlation function 
\eqn\IntNamp{
{\cal A}^{(N)}_p(k_1,\cdots,k_N) =\Int\, \prod_{i=1}^N d\x_I\,
\left\langle e^{ik^1\cdot{X}}(\xi_1)\cdots e^{ik^N\cdot{X}}(\xi_N)
\right\rangle_B }
is invariant if 
\eqn\invsgn{
\sgn(\x'_I-\x'_J)=\sgn({\x}_I-{\x}_J),}
is true. In the usual case of the real strings, this is true under
SL(2,$\Re$) transformations which preserve the cyclic ordering of
$\{\xi_1,\xi_2,\cdots,\xi_N\}$. In the $p$-adic case, the absence 
of an order compatible with the algebraic properties of $\Qp$, 
prevents us from making an analogous restriction. We will therefore
have to be content with a limited invariance. Namely, we will prove 
Eq.\invsgn\ for an infinitesimal transformation: 
\eqn\infMoebius{
\xi_I ~\to~ \xi'_I=\xi_I+\epsilon_{-1}+\epsilon_0\xi_I+
\epsilon_1\xi_I^2, }
which we require to satisfy
\eqn\infinitesimalcondition{
\abp{\epsilon_0}<1, \qquad \abp{\epsilon_1}\cdot\abp{\xi_I}<1,
\qquad (I= 1,2,\cdots,N).}
Therefore, $\sgn(\x'_I-\x'_J)=\sgn\l(\x_I-\x_J\r)\,\sgn\l(1+\ep_0+
\ep_1(\x_I+\x_J)\r)$. Thanks to the non-archimedian property: 
$\abp{\x_i+\x_j}\leq\max(\abp{\x_i},\abp{\x_j})$ and the condition
\infinitesimalcondition, 
$$
1+\ep_0+\ep_1(\x_I+\x_J) \in 1+p\,\Zp. 
$$
It can be shown that the sign function \defsgn\ of $p$-adic numbers of
this form is positive\GGPSP. This proves the invariance of the
integrated correlation function \IntNamp\ under infinitesimal
projective transformation. Unfortunately, however, this is not
sufficient to fix the positions of three of the vertex operators.


\newsec{Tachyon scattering amplitudes}

In this section, we evaluate the $N$ tachyon scattering amplitudes,
specifically for $N=3,4$. The lack of GL(2,$\Qp$) invariance due to
the $B$-field does not, strictly speaking, allow us to fix three of
the positions $\xi_I$. We will, nevertheless, fix $\xi_1$, $\xi_2$ and
$\xi_N$ by hand to $0,\,1$ and $\infty$. The last one requires some
care, as we can approach `infinity' through a sequence of `positive'
or `negative' $p$-adic numbers (see the discussion in Sec.~3.2). We
can average over the two, however, it turns out that the final result
is not sensitive to the details as long as the sequence of points have
the same value of $\sgn$.

Let us consider the three-tachyon amplitude to begin with. Ignoring
the momentum conserving delta function and an infinite factor, the
result is ${\cal A}^{(3)}_p(k_1,k_2,k_3)=e^{ {i\over2}(k_1\theta
k_2)}$. Since there is no natural notion of a cyclic (or indeed any)
ordering in $\Qp$, we need to add by hand a term with the role of
$k_1$ and $k_2$ interchanged and average over the two. The final
result is:
\eqn\pBtachthree{
{\cal A}_p^{(3)}(k_1,k_2,k_3) = 
\cos\half\!\left(k_{1}\theta k_{2}\right). }
This expression is the same as that of the usual bosonic string. 

Consider the case $N=4$ next. Once again, fixing $\xi_1,\,\xi_2$ and
$\xi_4$ by hand, we arrive at the integral:
\eqn\Bfourint{
e^{{i\over2}k_1\theta k_2}\,\int_{{\bf Q}_p}\!
d\xi\,|\xi|_p^{k_1\cdot k_3}
|1-\xi|_p^{k_2\cdot k_3}\,\exp\left({i\over2}(k_1\theta k_3)\sgn(\xi)
-{i\over2}(k_2\theta k_3)\sgn(1-\xi)\right).}
This can be evaluated in two ways: either we can expand the
exponential and use the expression of generalized $p$-adic
beta-/gamma-functions (given in Appendix A), or we can divide the
integral in four domains as in \BFOW. Following the first route, we
get, after averaging with a term with $(k_1\leftrightarrow k_2)$:
\eqn\Bfourbetagamma{
\eqalign{
{\cal A}_p^{(4)} =&\; c_{12}c_{13}c_{23}\,
\Gm(\a(s))\,\Gm(\a(t))\,\Gm(\a(u))
+ s_{12}s_{13}c_{23}\,\G(\a(s))\,\G(\a(t))\,\Gm(\a(u))\cr
&\,-s_{12}c_{13}s_{23}\,\G(\a(s))\,\Gm(\a(t))\,\G(\a(u))
+c_{12}s_{13}s_{23}\,\Gm(\a(s))\,\G(\a(t))\,\G(\a(u))\cr
=&\; c_{12}c_{34}\,
\l[{p-1\over{p}}{1\over{p}^{\a(s)}-1}-{1\over{p}}\r]
+ c_{13}c_{24}\,
\l[{p-1\over{p}}{1\over{p}^{\a(t)}-1}-{1\over{p}}\r]\cr
&\,+c_{14}c_{23}\,
\l[{p-1\over{p}}{1\over{p}^{\a(u)}-1}-{1\over{p}}\r]
+c_{12}c_{13}c_{23}\,{p+1\over{p}}, }}
where, we have used the abbreviations $c_{IJ}=\cos\half k_{I\mu}
\theta^{\mu\nu}k_{J\nu}$ and $s_{IJ}=\sin\half k_{I\mu}
\theta^{\mu\nu}k_{J\nu}$ and as usual, $\alpha(s)={s\over2}=
k_1\cdot{k_2}+1$, etc. Alternatively, following \BFOW, we divide $\Qp$
into $\Zp$ and its complement ${\cal D}_1$; then divide $\Zp$ further
into three domains: ${\cal D}_2$, in which the first coefficient
$\xi_0$ in the expansion of $\xi$ is $0$; ${\cal D}_3$ corresponding
to $\xi_0=1$ and ${\cal D}_4$ corresponding to the union of
$\xi_0=2,3,\cdots,p-1$.  In ${\cal D}_{1,2,3}$ the integral gives
${p-1\over p}{e^{{i\over2}k_1\theta k_2} c_{34}\over
p^{\alpha(s)}-1}$, ${p-1\over p}{e^{{i\over2}k_2\theta k_4}
c_{13}\over p^{\alpha(t)}-1}$ and ${p-1\over p}{e^{-{i\over2}k_1\theta
k_4} c_{23}\over p^{\alpha(u)}-1}$ respectively. In the region ${\cal
D}_4$, we are unable to do the calculation for an arbitrary prime $p$.
However, explicit evaluation for the first few relevant primes leads
to the result
\eqn\contact{ 
{e^{{i\over2}(k_1\theta k_2)}\over p}\left[(p-3)\, 
c_{13}c_{23}+e^{-{i\over2}(k_1\theta k_3)
+{i\over2}(k_2\theta k_3)}\right],} 
which, when averaged with the term with $(k_1\leftrightarrow k_2)$,
agrees with \Bfourbetagamma. Recall that in the commutative case, this
region (and likewise its analogues for higher $N$) gives the
four-tachyon (respectively $N$-tachyon) vertex in the spacetime
effective field theory, as the integrand here is independent of the
spacetime momenta in that case. With the $B$-field turned on, there is
always momentum dependence, but in this region, it is only in the
`phase' factors. Let us also note that the presence of the
$\Gamma_\tau$-function in the four-tachyon amplitudes makes it
impossible for these to have any adelic property.

Unfortunately, the four-point function is not symmetric under a
permutation of the four momenta. Even if we consider summing up the
terms with different permutations by hand, the resulting expression
does not match the results of the noncommutative deformation of the
tachyon effective field theory considered in \ncpadic. The latter is
a field theory with the action:
\eqn\NCTaction{
{\cal L}_{NC}^{p}(T) = {p\over p-1}\left[ 
-\half T\star p^{-\half\lform-1} T + \sum_{N=3}^{p+1}{(p-1)!\over
N!(p-N+1)!}\left({g\over p}\right)^N\,\left(\star T\right)^N
\right], }
where $T$ is the tachyon field on the D-brane and $g$ is the open 
string coupling. The three- and four-tachyon amplitudes from this
field theory are:
\eqn\FTresults{
\eqalign{
{\cal A}_{NC}^{p(3)} & \sim {p-1\over p}\, c_{12}\cr
{\cal A}_{NC}^{p(4)} & \sim {p-1\over p}\left(
{c_{12}c_{34}\over{p}^{\a(s)}-1} +
{c_{13}c_{24}\over{p}^{\a(t)}-1}
+ {c_{14}c_{23}\over{p}^{\a(u)}-1}\right)\cr
&\qquad + {p-2\over p}\,\left(c_{12}c_{34}+c_{13}c_{24}+c_{14}c_{23}
\right). }}
Finally, let us contrast the result \Bfourbetagamma\ with the 
four-tachyon amplitude in the usual noncommutative bosonic string 
theory:
\eqn\usualNCfour{
\eqalign{
A^{(4)}_{B} &= \cos\half(k_1\theta k_3-k_2\theta k_4)\,B(\a(s),\a(t))
+ \cos\half(k_1\theta k_4+k_2\theta k_3)\,B(\a(t),\a(u))\cr
&\qquad + \cos\half(k_1\theta k_2+k_3\theta k_4)\,B(\a(u),\a(s)),}}
where we have averaged over cyclic and anti-cyclic permutations and
$B(\alpha,\beta)$ is the beta function. 


\newsec{Alternative proposal for the four-tachyon amplitude}

In this section, we will explore the possibility of defining
four-tachyon amplitudes of $p$-adic string in a constant $B$-field so
as to reproduce the noncommutative deformation \NCTaction\ of the
effective field theory\ncpadic. To this end, let us recall the
situation for the usual bosonic string theory. Either in the presence
of the $B$-field or with the Chan-Paton factor, the coefficient of the
noncommuting factors contributing to ${\cal A}_B^{(4)}$ in the
different channels, can be summed up to write a single integral over
the real numbers\Fairlie. In \Barnett, which considers introducing the
Chan-Paton factor in $p$-adic strings, it was suggested that this
property be abandoned. Following this idea, we will consider different
channels separately.  Consider the $t$-$u$ channel first: in the
ordinary string theory, this is given by:
\eqn\tuVeneziano{
{\cal A}^{(4)}_{B,tu} = 
\int^1_0 d\xi\,|\xi|^{k_1\cdot{k}_3}|1-\xi|^{k_2\cdot{k}_3}
\cos\,\half\left(k_1\theta{k}_2\,+\,\epsilon(\xi)\,k_1\theta{k}_3\,
+\,\epsilon(1-\xi)\,k_2\theta{k}_3\right), }
where, $\epsilon(\xi)$ is the sign function on $\Re$. We will attempt
to write an appropriate $p$-adic analogue of the above. However, there
does not exist a unique prescription for this, a fact that was already
realized in \refs{\Heaviside,\Barnett} in the context of the Chan-Paton
factors. In order to generalize \tuVeneziano\ to the $p$-adic case, we
need to insert a suitable projector to restrict the range of
integration.  There are several possibilities\foot{Our choice,
however, is somewhat more restricted by the fact that we consider only
$\t=p,\,p\ve$ with prime $p\equiv3$ ($\mod 4$), so as to ensure
$\sgn(-\xi)=-\sgn(\xi)$. This limits our options compared to
\Heaviside.} to achieve this, as has been listed by the authors of 
\Heaviside:
\eqn\steps{
\eqalign{
\Xi_1(\x) &= {1\over4}\left[1\,+\,\sgn(\x)\right]\,\left[1\,+\,
\sgn(1-\x)\right]\,\equiv\,
H_\tau(\xi)\,H_\tau(1-\xi),\cr
\Xi_2(\x) &= \hf\l[1\,+\,\sgn(\x)\,\sgn(1-\x)\r],\cr
\Xi_3(\x) &= \hf\l[\sgn(\x)\,+\,\sgn(1-\x)\r],}}
where, $H_\tau(\xi)=\half\left(1+\sgn(\xi)\right)$ is the $p$-adic
analogue of the Heaviside function. 
Inserting one of the step functions above, we can thus define:
\eqn\ptu{
\eqalign{
{\cal A}^{(4)}_{tu,\Xi} &=
\Int d\xi\;\Xi(\xi)\,{\abp{\x}^{k_1\cdot{k}_3}
\abp{1-\x}^{k_2\cdot{k}_3}}\cr
&\qquad\times
\cos\,\half\l( \sgn(-1)\,k_1\theta{k}_2\,+\,\sgn(-\x)\,
k_1\theta{k}_3\,+\,\sgn(1-\x)\,k_2\theta{k}_3\right). }}
In the following, we will study the different choices in turn and 
examine whether a field theoretic interpretation is possible in 
each case.

Let us consider $\Xi_1$ first.
\eqn\ptuone{
\eqalign{
{\cal A}^{(4)}_{tu,\Xi_1} 
&={1\over4}\cos{k_1\theta{k}_4 + k_2\theta{k}_3\over 2}
\bigg[\,\Gm(\a(s))\,\Gm(\a(t))\,\Gm(\a(u)) - 
\G(\a(s))\,\G(\a(t))\,\Gm(\a(u))\cr
&\quad - \G(\a(s))\,\Gm(\a(t))\,\G(\a(u))
+\Gm(\a(s))\,\G(\a(t))\,\G(\a(u))\bigg]\cr
&= \hf\cos {k_1\theta{k}_4 + k_2\theta{k}_3\over2}\,
\l[{p-3\over2\,p}+{p-1\over p}\l\{
{1\over p^{\a(u)}-1}+{1\over p^{\a(t)}-1}\r\}\r]. }}
Upon summing over the three, namely, $t$-$u$, $u$-$s$ and $s$-$t$
channels, we find that the resulting Veneziano amplitude agrees
qualitatively with the calculation from the effective field theory
\NCTaction.

For the other choices of the step function in \ptu, the results are:
\eqn\otherptu{
\eqalign{
{\cal A}^{(4)}_{tu,\Xi_2} 
&= c_{12}\left(c_{13}c_{23}+s_{13}s_{23}\right)\,
{\cal B}^{(4)}_{tu,\Xi_2}\,
-\, s_{12}\left(s_{13}c_{23}-c_{13}s_{23}\right)\,
{\cal B}^{(4)}_{tu,\Xi_3}\cr
{\cal A}^{(4)}_{tu,\Xi_3} 
&= c_{12}\left(c_{13}c_{23}+s_{13}s_{23}\right)\,
{\cal B}^{(4)}_{tu,\Xi_3}\,
-\, s_{12}\left(s_{13}c_{23}-c_{13}s_{23}\right)\,
{\cal B}^{(4)}_{tu,\Xi_2}, }}
where,
\eqn\Btu{
{\cal B}^{(4)}_{tu,\Xi} =
\Int d\xi\;\Xi(\xi)\,{\abp{\x}^{k_1\cdot{k}_3}
\abp{1-\x}^{k_2\cdot{k}_3}}. }
Explicitly, 
\eqn\Btuonetwo{
\eqalign{
{\cal B}^{(4)}_{tu,\Xi_2} &=\,
\hf\bigg[\Gm(\a(s))\,\Gm(\a(t))\,\Gm(\a(u))
+\Gm(\a(s))\,\G(\a(t))\,\G(\a(u))\bigg]\cr
&=\; {p-1\over2{p}}\l(1+ {1\over{p}^{\a(t)}-1}+{1\over{p}^{\a(u)}-1}
\r),\cr
{\cal B}^{(4)}_{tu,\Xi_3} &=\,
-\,\hf\bigg[\G(\a(s))\,\Gm(\a(t))\,\G(\a(u))
+\G(\a(s))\,\G(\a(t))\,\Gm(\a(u))\bigg]\cr
&=\; -{1\over{p}}+{p-1\over2{p}}
\l({1\over{p}^{\a(t)}-1}+{1\over{p}^{\a(u)}-1}\r). }}
Hence:
\eqn\othertwo{
\eqalign{
{\cal A}^{(4)}_{tu,\Xi_{2,3}} &=\,
\cos\hf\l(k_1\theta{k_4}+k_2\theta{k_3}\r)\,
\l[{p-3\over{4p}}+{p-1\over2{p}}
\l({1\over{p}^{\a(t)}-1}+{1\over{p}^{\a(u)}-1}\r)\r]\cr
&\qquad \pm\,\cos\hf\l(k_1\theta{k_2}+k_2\theta{k_3}+k_3\theta{k_1}\r)\,
{p+1\over4p}, }}
the two choices differ by the sign of the term in the second line. When 
we sum over the three channels, the result does not match that obtained
from a field theory---it is the second term which is not compatible.
Thus, among the step functions in \steps, only  $\Xi_1(\x)$ allows for
a field-theoretical interpretation. There is one drawback of this
prescription, however. Upon turning off the background $B$-field, 
the amplitude does not reduce to the original amplitude. 
The difficulty here is reminiscent of the same problem encountered in
introducing Chan-Paton factors. It is likely that the source of the
problem and possibly its solution have the same origin. 

Formally, this prescription can be extended to higher-point amplitudes 
in one of the channels by insertion of the Heaviside functions as:
\eqn\higherH{
\eqalign{
&\Int d\x_3\cdots d\xi_{N-1}\prod^{N-1}_{I=3}
\abp{\xi_I}^{k_1\cdot{k}_I}\abp{1-\xi_I}^{k_2\cdot{k}_I}
H_\tau(\xi_I)\,H_\tau(1-\xi_I)\cr
&\;\times\prod^{N-1}_{\buildrel{I,J=3}\over{I<J}}
\abp{\xi_I-\xi_J}^{k_J\cdot{k}_J}\,H_\tau(\xi_J-\x_I)
\,\exp\l\{-{i\over2}\sum^{N-1}_{\buildrel{I,J=1}\over{I<J}}
(k_I\theta k_J)\,\sgn(\xi_I-\xi_J)\r\}. }}
Summing over all the permutations under the exchange of momenta of the
above amplitude is expected to yield full $N$-point amplitude. However, 
we will not attempt to check if this agrees with the corresponding
amplitude derived from the effective field theory.

\newsec{Discussions}

In summary, we have proposed a coupling for the antisymmetric tensor
field $B$ in the $p$-adic string theory. Specifically, we have added a
term corresponding to a constant $B$-field to the nonlocal action on
the boundary $\Qp$ of the $p$-adic `worldsheet'. The exact Green's
function for the fields $X^\mu(\xi)$ is computed in the presence of
the constant $B$-field. The results parallel the case of the usual
string theory in a constant $B$-field\refs{\Schom,\SW}: there is a
noncommutative factor in the correlation functions of the tachyons and
the flat `closed string' metric is replaced by the open string metric
$G_{\m\n}$. However, the action with the $B$-field is not invariant
under M\"obius transformation, a problem that is intimately connected
with the fact that there is no natural order among the $p$-adic
numbers. An unfortunate consequence is that the resulting tachyon
amplitudes do not quite match those obtained from the noncommutative
deformation of the tachyon effective field theory.  We also examined
the possibility of defining Koba-Nielsen amplitudes so as to derive
the field theory results.

Our objective was to derive the noncommutative deformation of the
spacetime effective action of the $p$-adic tachyon, proposed in
Ref.\refs{\moscow,\ncpadic}, from a `worldsheet' point of view.  In
spite of having some encouraging results, this problem remains
unresolved. We would expect that for a proper understanding of the
$B$-field, one has to deal with {\it closed} $p$-adic strings of which
virtually nothing is known. In the early days, in analogy with the
usual strings, it was thought that the closed strings have to do with
the quadratic extension of $\Qp$. Unlike $\Re$, however, neither is
the quadratic extension unique, nor is it closed. Indeed, there are an
infinite number of finite extensions of $\Qp$ and none of these is a
closed field. Moreover, each finite extension is isomorphic
to the boundary of a tree (a Bethe lattice) with appropriate
coordination number\Zabro. Thus, they would correspond to some kind of
generalized open strings. It is possible to define a closed field by
taking the union of these extensions and augmenting the set by putting
in limit points of sequences. The resulting field ${\bf C}_p$,
surprisingly, is isomorphic to the set of complex numbers\RobP,
although their topologies are very different. It is likely that 
closed $p$-adic strings are to be based on this field ${\bf C}_p$. 

Be that as it may, a constant $B$-field is certainly simpler. First,
it only affects the boundary terms. Secondly, one can think of the
open $p$-adic `worldsheet', at least for $p=2$, as an exotic
discretization of the upper half-plane with its Poincare
metric\NeVa. Thus, one should be able to capture the constant flux of
the $B$-field at the vertices of the tree. Finally, let us comment on
how one may proceed to couple the $B$-field to the bulk `worldsheet'
action. To this end, we note that the effect of the  $B$-field
is topological. We add the pull-back of the 2-form: $X_*(B)=
B_{\mu\nu}\del_\alpha X^\mu\del_\beta X^\nu\epsilon^{\alpha\beta}$ to
the worldsheet action. However, the `worldsheet' of the $p$-adic
string, being a tree, has no closed loop. Hence naively, there is no
2-cycle over which to integrate a 2-form. Nevertheless, we notice that
for a prime $p$ (or for any odd integer for that matter), the tree is
a bipartite one. We can consistently divide the edges $E=\{e\}$ into
two disjoint sets $E_1=\{e_1\}$ and $E_2=\{e_2\}$. It is now possible
to antisymmetrize derivatives along the edges belonging to these two
subsets:
\eqn\bulkB{
B_{\mu\nu}\,\Delta_{\left[e_1\right.}X^\mu(z)\;
\Delta_{\left. e_2\right]}X^\nu(z). }
Note that this division does not correspond to the $\sgn$ function. 
While the proposal above is unlikely to work as it is, it may be 
interesting to ponder along these lines. 

\bigskip

\noindent{\bf Note added:} A paper \Grange\ (hep-th/0409305) 
suggesting the same way to couple the $B$-field in the $p$-adic 
string as that considered here, appeared on the arXiv as this 
manuscript was being finalized. The conclusions there are formal
and suffer from subtleties that we have discussed in Sec.~3.

\bigskip

\centerline{\bf Acknowledgement} 
\noindent We are grateful to Chandan Singh Dalawat and Seiji Terashima
for useful discussion.  DG would like to thank Tohru Eguchi and the
Japan Society for the Promotion of Science (JSPS) for an Invitation
Fellowship which partially supports this work. The research of TK is
supported in part by the Grants-in-Aid (\#16740133) and (\#16081206)
from the Ministry of Education, Science, Sports and Culture of Japan.


\appendix{A}{Materia $p$-adica}

In this appendix, we provide a compendium of formulas related to the
$p$-adic number field $\Qp$ and functions over it. It is not intended 
as a review, but only to collect in one place most of what is used in 
this article. For details, we refer the reader to 
Refs.\refs{\GGPSP\KobP\RobP-\RussP}. 

\subsec{$p$-adic numbers and the Bruhat-Tits tree}

An element $\xi\in\Qp$ may be written as a power series in $p$: 
\eqn\plaurent{
\xi = p^{N}\l(\xi_0+\xi_1p+\xi_2p^2+\cdots\r)
= p^{N}\sum_{n=0}^\infty\xi_n p^n, }
where $N\in\ZZ$ and $\xi_n\in\{0,1,\cdots,p-1\}$, $\xi_0\ne 0$.
(There is, however, nothing special about this choice---one may work
with other representative elements.)  Notice the similarity with the
Laurent series of meromorphic functions. Addition and multiplication
in $\Qp$, defined in terms of the series \plaurent, give $\Qp$ the
structure of a ring. The subset $\Zp$ consisting of elements with
$N\geq 0$ in \plaurent\ is a subring known as the $p$-adic
integers. The non-archimedian $p$-adic norm of $\xi$ is defined to be
\eqn\pnorm{
\abp{\xi}=p^{-N}, \qquad \abp{0}=0; }
which satisfies the inequality 
$\abp{\xi-\xi'} \leq \max\l(\abp{\xi},\abp{\xi'}\r)$,
which is stronger than the usual triangle inequality. Notice that the
norm of a $p$-adic integer is at most 1. 

The series representation \plaurent\ of a $p$-adic number also
provides us with an isomorphism between $\Qp$ and the boundary
$\del{\cal T}_p$.  To see this, let us consider the case $p=3$ for
definiteness and refer to Fig.~1, in which the dashed line denotes the
(unique) path through the tree connecting $\xi=\infty$ to $\xi=0$.
Let us label the vertices along this path by $C_N$, with $N=\pm\infty$
corresponding to zero and infinity respectively and $N=0$ being the
(arbitrarily chosen) point $C=C_0$ at the `centre' of the tree.  In
order to find a point on the boundary $\del{\cal T}_p$ corresponding
to $\xi$, we start at $C_N$ and then choose the left branch if
$\xi_0=1$ or the right one if $\xi_0=2$. At the next step, we choose
one of the branches depending on the value of $\xi_1$, and so
on. Continuing this way, we arrive at a point on the boundary.  This
procedure can obviously be generalized for any $p$.

Coming back to the series \plaurent, the finite part
$\displaystyle\sum_{n=0}^{N-1}\xi_np^{n-N}$, consisting of negative
powers of $p$, is called the {\it fractional part} $\{\xi\}_p$ of
$\xi$. The rest, an infinite series in general, is the {\it integer
part} $[\xi]_p$.  Thinking of the fractional part $\{\xi\}_p$ as real,
let us define the complex valued function $\ch:\Qp\to\CC$, as
\eqn\pexponential{
\ch(\xi) = \exp\left(2\pi i\{\xi\}_p\right)
=\exp\left(2\pi i\xi\right), }
where, in writing the second expression, we have allowed for a little
imprecision to regard \plaurent\ as a formal power series in real.  As
in the real case, the contribution to \pexponential\ from the integer
part $[\xi]_p$ is trivial and only the fractional part matters. Since
$\ch(\xi+\xi')=\ch(\xi)\ch(\xi')$, it is called an {\it additive
character} of $\Qp$. 

\subsec{Fourier transformation}

The Fourier transform of a complex valued
function\foot{Strictly speaking, some of the proofs hold directly for
a locally constant function $f(\xi)$, which is a function such that
for any $\xi\in\Qp$, $f(\xi)=f(\xi+\xi')$, whenever
$\abp{\xi'}\leq{p}^{l(\xi)}$, for some integer $l(\xi)\in\ZZ$. 
The results are then extended to the generalized functions, 
which are defined as continuous functionals on the linear vector
space of locally constant functions. 

For any locally constant function $f(\xi)$ with compact support, there
exists a maximum value $l(f)=\max_{\xi\in\Qp}l(\xi)$, called the
parameter of constancy.  Let $\cD_m^l$ be the set of locally constant
functions with support contained in
$B_m=\left\{\xi\in\Qp\;|~\abp{\xi}\leq p^m\right\}$ and with the
parameter of constancy at least $l$: $\cD_m^l=\l\{f\in\cD\;|~{\rm supp
}f\subset{B}_m,~ l(f)\geq{l}\right\}$. The Fourier transform $\wt
f(\omega)$ of $f(\xi)\in\cD_m^l$ turns out to be in $\cD_{-l}^{-m}$.}
is defined by:
\eqn\FT{
\wt f(\omega)=\Int d\x~ \ch\l(-\omega \xi\r) f(\xi), }
where $d\xi$ is the translationally invariant Haar measure on $\Qp$
normalized as $\IntZ d\xi=1$. The inverse Fourier transformation is
given by
\eqn\invFT{
f(\xi)=\Int d\omega\,\ch\l(\omega\xi\r)\wt f(\omega). }
This can be proven by use of the formula
\eqn\intexp{
\int_{B_N}\ch\l(\omega\xi\r)d\xi=
\cases{p^N &for $\abp{\omega}\leq{p}^{-N}$,\cr
0 &for $\abp{\omega}>{p}^{-N+1}$}, }
however, we will omit the proof. It is useful to have the following
representation of the delta function $\delta(\xi)$: 
\eqn\FTdel{
\delta(\xi)=\Int d\omega\, \ch(\omega{\xi}). }
It gives $\Int d\xi\,f(\xi)\,\delta(\xi)= f(0)$ as expected. 

\subsec{Generalized $p$-adic gamma and beta functions}

The usual $p$-adic gamma function, called the Gelfand-Graev-Tate 
gamma function, is defined as:
\eqn\tGGgamma{
\Gamma(s) = \Int d\xi\, \ch(\xi)\,|\xi|_p^{s-1}={1-p^{s-1}\over 
1-p^{-s}}. }
It has only one singular point at $s=0$, a simple pole with residue 
$(p-1)/p\ln p$. The generalized gamma function associated with 
$\sgn$ is:
\eqn\genGamma{
\eqalign{
\Gamma_\tau(s) &= \Int d\xi\, \ch(\xi)\,|\xi|_p^{s-1}\sgn(\xi)\cr
&=\pm\sqrt{\sgn(-1)}\,p^{s-\half}\quad\hbox{\rm for } p=p,\;
\varepsilon p. }}
This is an entire function with no singularity on the $s$-plane.
Using the above, we can define the following generalized $p$-adic
Beta functions:
\eqn\pBeta{
\eqalign{
\Int d\x\; |\xi|_p^{x-1} |1-\xi|_p^{y-1}
&=\Gm(x)\,\Gm(y)\,\Gm(1-x-y)\cr
\Int d\x\; \sgn(\x)\,|\xi|_p^{x-1} |1-\xi|_p^{y-1}
&=\sgn(-1)\,\Gm(x)\,\G(y)\,\G(1-x-y),\cr
\Int d\x\; \sgn(1-\x)\,|\xi|_p^{x-1} |1-\xi|_p^{y-1}
&=\sgn(-1)\,\G(x)\,\Gm(y)\,\Gm(1-x-y),\cr
\Int d\x\; \sgn(\x)\,\sgn(1-\x)\,|\xi|_p^{x-1} |1-\xi|_p^{y-1}
&=\G(x)\,\G(y)\,\Gm(1-x-y). }}

\subsec{Fourier transform of derivatives}

We will now discuss the Fourier transform of the derivatives of a
(complex valued) function on $\Qp$ or its extension to the interior of
the `worldsheet' ${\cal T}_p$. Recall the expressions for the
normal and tangential derivatives in Eqs.\normal\ and \tangent.
More generally, the $\ell$-th normal derivative is given by
\eqn\ellnormal{
\l(\del^{(p)}_n\r)^\ell f(\xi) = \Int d\xi'\, {f(\xi')-f(\xi)\over
|\xi'-\xi|_p^{\ell+1}}. }
After a Fourier transform, we obtain
\eqn\FTnormalone{
\l(\del^{(p)}_n\r)^\ell f(\xi) = \Int d\omega\, \ch(\omega{\xi})
\l(\widetilde{\del^{(p)\ell}_n f}\r) (\omega), }
where
\eqn\FTnormaltwo{
\l(\widetilde{\del^{(p)\ell}_n f}\r) (\omega) =\wt f(\omega)
\Int d\xi\,{\ch(-\omega{\xi})-1\over|\xi|_p^{\ell+1}}. }
The integral above is divergent. Therefore, in order to evaluate it,
we need to regularize by an `infrared' cutoff. This is done as
follows. Let $S_m=\{\xi\in\Qp\;|~\abp{\xi}=p^m\}$, the subset of
$p$-adic numbers with a fixed norm $p^m$. Clearly, $S_m\cap
S_{n}=\emptyset$ for $m\ne n$, and $\Qp=\cup_{m\in{\bf Z}}S_m$. We define
\eqn\regint{
\eqalign{
\Int {d\xi\over|\xi|_p^{\ell+1}}
&=\lim_{N\to\infty}\sum^\infty_{m=-N}\int_{S_m}{d\xi\over
|\xi|_p^{\ell+1}}\cr
&={p-1\over p}\,\lim_{N\to\infty}\sum^N_{m=-\infty}p^{m\ell},\cr
\Int d\xi\,{\ch(\omega{\xi})\over|\xi|_p^{\ell+1}}
&=\lim_{N\to\infty}\sum^\infty_{m=-N}\int_{S_m}d\xi
\,{\ch(\omega{\xi})\over|\xi|_p^{\ell+1}}\cr
&=\lim_{N\to\infty}\l(\sum^N_{m=k}p^{m\ell}-\sum^N_{m=k-1}
p^{m\ell-1}\r)\cr
&=\lim_{N\to\infty}\left[{p-1\over p}\sum^N_{m=k} p^{m\ell} -
p^{(k-1)\ell-1}\right], }}
where $k\in\IZ$ is an integer such that $\abp{\omega}=p^k$. In the above,
we have used the fact that in our normalization $\int_{S_m} d\xi = 
(p-1)p^{m-1}$. Thus we obtain 
\eqn\regFTnormal{
\Int d\xi\,{\ch(-\omega{\xi})-1\over|\xi|_p^{\ell+1}}
=|\om|_p^\ell\;{1-p^{-\ell-1}\over1-p^\ell}, }
specifically for $\ell=1$: 
\eqn\FTnder{
\l(\widetilde{\del^{(p)}_n f}\r) (\omega)
={1-p^{-2}\over1-p}\,\abp{\omega}\,\wt f(\omega). } 

Finally let us turn to the Fourier transform of the tangential
derivative. After a change of the variable in \genGamma, one finds 
\eqn\FTsgn{
{\sgn(-\omega)\over|\om|_p^s}\,\Gamma_\tau(s)
=\Int d\xi\, \ch(-\omega\xi)\,|\xi|_p^{s-1}\sgn(\x) .}
Using the above, it is easy to find the Fourier transform of the 
tangential derivative:
\eqn\FTtder{
\eqalign{
\l(\widetilde{\del^{(p)}_t f}\r) (\omega)
&=\Int d\xi d\xi'\, {\sgn(\xi-\xi')\over|\xi-\xi'|_p^2}\,
\ch(-\omega\xi) f(\xi')\cr
&=\Gamma_\tau(-1)\,\sgn(-\omega)\,\abp{\om}\,\wt f(\omega). }}
Let us reiterate that $\Gamma_\tau(-1)$ is not singular.

\appendix{B}{A regularized integral}

In this appendix, we provide a brief derivation of an integral that
was used in the main part of the text in finding the solution \solnX\
of the `equation of motion', or equivalently, the Green's function 
\Greensol. The integral in question is
\eqn\twopoint{
\Int d\omega\, {\ch(\omega{\xi})-\ch(\omega{\xi'})\over\abp{\omega}}
=-\,{p-1\over{p}\ln{p}}\ln{\abp{\x}\over\abp{\x'}}, }
which follows from the `infrared regularized' expression
\eqn\propagator{
\Int d\omega\, {\ch(\omega{\xi})\over\abp{\omega}} =
{p-1\over{p}\ln{p}}\l[\lim_{\a\to0}\l({1\over\a}\r)-\ln\abp{\x}\r].}
In order to obtain this, we consider the following more general integral,
well-defined for $\alpha>0$, and evaluate it as: 
\eqn\IRreg{
\eqalign{
\Int |\omega|_p^{\alpha-1}\,\ch(\omega\xi)\,d\omega
&=\sum^{\infty}_{m=-\infty}\int_{S_m} |\omega|_p^{\alpha-1}\,
\ch(\omega\xi)\,d\omega\cr
&={1\over|\omega|_p^{\alpha}}\l({1-p^{\alpha-1}\over 
1-p^{-\alpha}}\r), }}
where $k\in\ZZ$ is an integer such that $\abp{\omega}=p^k$. The
reguralized expression \propagator\ is given in the limit
$\alpha\to0$.  The procedure here exactly parallels the case of
standard two-dimensional theories, such as the one for the worldsheet
theory of the usual strings.


\listrefs 

\bye